\documentclass[a4paper,floatfix,rmp,twocolumn,showkeys,superscriptaddress]{revtex4}
  \usepackage[latin1]{inputenc}
  \usepackage{graphicx}
  \usepackage{mathtools}
  \setcitestyle{numbers,square,comma}
  \usepackage{amsmath}
  \usepackage{makecell}
  \usepackage{color}
  \usepackage{hyperref}

\begin{document}

\title{Hierarchical genotype networks and incipient ecological speciation in Q$\beta$ phage quasispecies}

  \providecommand{\ASU}{Center for Fundamental and Applied Microbiomics and School of Life Sciences, Arizona State University, Tempe, AZ 85287, USA}
  \providecommand{\CNB}{Departamento de Biolog\'ia de Sistemas, Centro Nacional de Biotecnolog\'ia (CSIC), C/ Darwin 3, 28049 Madrid, Spain. }
  \providecommand{\GISC}{Grupo Interdisciplinar de Sistemas Complejos (GISC), Madrid, Spain. }
  \providecommand{\CAB}{Department of Molecular Evolution, Centro de Astrobiolog\'ia (CSIC), Instituto Nacional de T\'ecnica Aeroespacial (INTA), Carretera de Ajalvir, Km 4, Torrej\'on de Ardoz, 28850 Madrid, Spain. }
  
 \author{Lu\'is F Seoane}
    \affiliation{\CNB}
    \affiliation{\GISC}

  \author{Henry Secaira-Morocho}
    \affiliation{\ASU}
  
  \author{Ester L\'azaro}
    \affiliation{\CAB}
  
  \author{Susanna Manrubia}
    \affiliation{\CNB}
    \affiliation{\GISC}

  \vspace{0.4 cm}
  \begin{abstract}
    \vspace{0.2 cm}

    Understanding how viral mutant spectra organize and explore genotype space is essential for unraveling the mechanisms driving evolution at the finest scale. Here we use deep-sequencing data of an amplicon in the A2 protein of the RNA bacteriophage Q$\beta$ to reconstruct genotype networks with tens of thousands of different haplotypes. The study of populations evolved under different temperature regimes uncovers generic topological features conditioned by fundamental structural motifs of genotype networks---tetrahedrons, triangles, and squares---that govern their local architecture. Mutant swarms display a hierarchical structure where sequences cluster around a highly connected and abundant sequence core that sustains population diversity. The immediate neighborhood of this core is comprehensively sampled, with no signs of selection, while a few mutations away sampling becomes dynamical and sparse, showing signs of purifying selection. By aggregating genotype networks from populations adapted to different temperatures, we capture the early stages of evolutionary divergence, with overlapping populations that remain connected through short mutational paths. Even at the time scale of these experiments, evolutionary pathways might be multiple, preventing the backward reconstruction of unique trajectories once mutations have been fixed. This analysis provides a detailed view of the local, fine-scale processes shaping viral quasispecies evolution and underscores the usefulness of genotype networks as an enlightening visualization of the organization of mutant swarms.

  \end{abstract}

  \keywords{Genotype network, quasispecies, topology}

\maketitle

	Visual metaphors are powerful and enlightening representations of biological concepts. They have been instrumental in the description, formalization and understanding of evolutionary processes, often sustaining the conceptual framework on which qualitative and quantitative advances rely. One of the most iconic images is the double helix, which captures at once the aesthetic beauty of the DNA molecule and its dual role as stable repository of genetic information and template for replication \cite{watson:1953}. The first depiction of the Tree of Life in Darwin's notes brought similar awareness, in this case to sort and relate organisms over a bigger picture \cite{darwin:1859}. Much later, once the main mechanisms behind such trees were understood, systematic reconstructions could be made, leading to a revolution in our understanding of life---e.g., as ribosomal RNA revealed the existence of Archaea \cite{woese:1977}. A similar clarity came from the description of the evolutionary process as the movement of populations on landscapes \cite{wright:1932}, where valleys and peaks represented dysfunctional or highly adapted states, respectively. 

	Powerful images underlie strong metaphors, but may also introduce notable limitations, since they condition the way in which we think about biological processes \cite{pigliucci:2012,olson:2019} and our interpretation of empirical observations. Metaphors, therefore, need to be critically examined, revised and updated alongside scientific advances. Phylogenetic trees and adaptive landscapes, specifically, have provided a broadly used framework to interpret evolutionary dynamics for about a century \cite{svensson:2012}. Since the advent of sequencing techniques, however, the idea of representing evolutionary dynamics as displacements on networks of neutral or quasi-neutral genotypes mutually attainable through unit mutational steps \cite{maynard-smith:1970} has been explored in various synthetic \cite{gruner:1996b,bastolla:2003} and empirical systems \cite{schultes:2000,payne:2014b,williams:2022}. This change of viewpoint has entailed changes in suitable representations of adaptive landscapes \cite{catalan:2017} and affected our understanding of adaptive dynamics \cite{schaper:2014,vahdati:2016,greenbury:2022,catalan:2023}. Analyses of the architecture of complete genotype networks in simple genotype-phenotype models have uncovered the existence of generic structural properties relevant for molecular evolution \cite{aguirre:2011,garcia-martin:2018,greenbury:2015,ahnert:2017} that need to be incorporated into evolutionary theory \cite{manrubia:2021}. For example, genotype networks have cycles that reveal the non-uniqueness of evolutionary trajectories  \cite{wagner:2014b,vahdati:2016,williams:2022,dabilla:2024}. In consequence, the number of mutations separating two individuals does not necessarily imply a temporal order of mutation fixation, but  potential pathways that enable bidirectional evolution from each other. This structural feature cannot be included in tree-like representations which, by construction, are acyclic graphs.

	Viral populations are large and heterogeneous ensembles of related genomic sequences \cite{domingo:2006,andino:2015} whose description and analysis are relevant in studies of pathogenesis and viral adaptability \cite{domingo:2021}, but also highly useful to dive into the intricacies of molecular evolution \cite{manrubia:2006}. RNA viruses are particularly prone at generating variation, with mutation rates that vary between $10^{-4}$ and $10^{-6}$ per nucleotide copied \cite{drake:1999,sanjuan:2010}. Mutant swarms not only function as repositories of variation, since multiple interactions among genomes in the quasispecies reveal a rich social life \cite{leeks:2023} and turn the quasispecies into a unit of selection. Significant efforts have been devoted to analyzing and visualizing the organization of viral quasispecies \cite{baccam:2001,henningsson:2019,delgado:2021,dabilla:2024}, as the way a mutant swarm is structured is key to understanding the role of diversity in adaptation \cite{somovilla:2019,delgado:2024} or to predicting emerging phenotypes \cite{skums:2017}, among others. Reconstruction of full viral genomes within the quasispecies is difficult due to their high variability and the limited read lengths in high-throughput sequencing \cite{baaijens:2017,ahn:2018}. For this reason, many studies focused on quasispecies diversity often use data from amplicon reads, permitting the exploration of ensembles with hundreds to a few thousand different haplotypes. This approach has allowed to infer fitness landscapes in HIV-1 \cite{lorenzo-redondo:2014}, to characterize temporal dynamics of sequences \cite{henningsson:2019}, to quantify the organization of the mutant spectrum in HCV \cite{delgado:2021}, or to identify functional diversification in SARS-CoV-2 \cite{delgado:2024}. 

	In this contribution, we reconstruct genotype networks with tens of thousands of different haplotypes obtained from the deep sequencing of an amplicon of the A2 protein of the RNA bacteriophage Q$\beta$ evolving under various environmental conditions. We show that the structure of these empirical genotype networks results from simple motifs that emerge from fundamental mathematical properties of sequence spaces. Mutant swarms comprehensively sample these motifs around a network center, generating a hierarchical structure. The exploration of biological possibilities becomes constrained by the genotype network topology, which determines accessibility between genotypes, and by an implicit fitness landscape, which biases the abundance of different genotypes but does not significantly limit the exploration of nearby mutants. Proper visualization of these large networks is aided by two algorithms specifically devised to create representations that emphasize the inherent organization of the reconstructed networks and, therefore, of viral quasispecies. The large size of the reconstructed networks, in turn, allows to perform statistically reliable measures of their structure and to formally derive topological features that should be generic in empirical genotype networks at various evolutionary scales. Finally, by merging three networks of Q$\beta$ for different environmental conditions, a synthetical picture of incipient speciation at the molecular level emerges, showing at once reversibility at short time scales and the blurry origin of phylogenetic branches.

	\section*{Results}

		\subsection*{Empirical data and genotype network reconstruction}
			\label{sec:gNets}
    
\begin{figure*}[h]
				\centering
				\includegraphics[width=\textwidth]{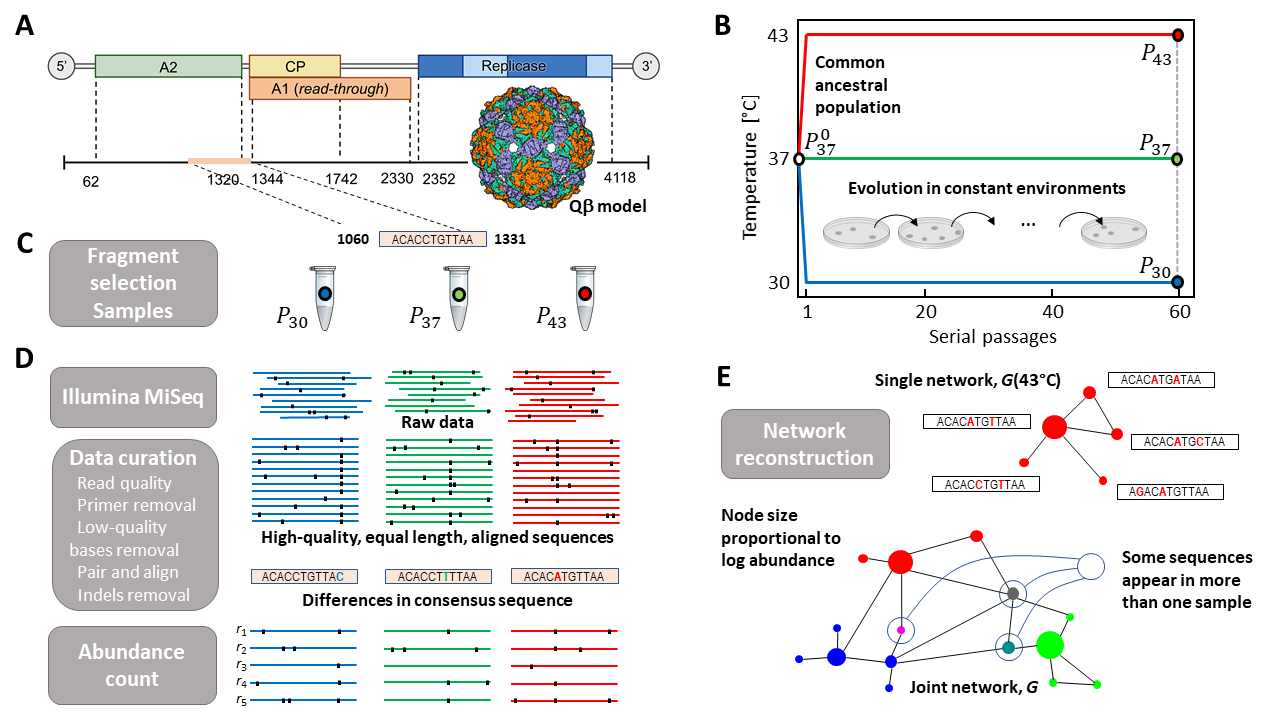}
				\caption{{\bf Schematic of evolution experiments, data processing and network reconstruction.} A. The Q$\beta$ phage has a linear, single-stranded RNA genome of positive polarity with a length of 4217 nucleotides. It is composed of four proteins: a maturation-lysis protein (A2), a capsid protein (CP), a read-through protein (A1), and an RNA-dependent RNA-polymerase (Replicase). A computational model of the icosahedral capsid is shown. B. Three replicates of an ancestral population pre-adapted to 37$^\circ$C for two passages, $P^0_{37}$ have been propagated for 60 passages at three different temperatures (see details in Materials and Methods). C. A lineage of the three replicas evolved in each condition was deep sequenced. A fragment with a length of 271 nt belonging to the A2 protein and a small non-coding region was selected. D. Sequencing yielded single-genome fragments (raw data) that were curated through a devoted pipeline, eventually returning populations with about $2-3 \times 10^5$ high-quality, equal length, aligned reads. We observed differences in the consensus sequence of each evolved population. Each set was scanned for repeated sequences through an abundance count algorithm. E. Individual and aggregated genotype networks were reconstructed. Each sequence corresponds to a node of the network, with size proportional to the logarithm of the abundance of the sequence; two nodes are connected through a link if they differ in a single point mutation.} 
				\label{fig:experiment}
			\end{figure*}

			Figure~\ref{fig:experiment} depicts a summary of the experiments and protocols carried out to reconstruct genotype networks for evolved populations of the Q$\beta$ phage. This phage has a short RNA genome (4217 nt) that encodes four proteins, Figure~\ref{fig:experiment}A. It infects the bacterium {\it Escherichia coli} using as receptor the conjugative F pilus. Its optimal replication temperature is 37$^{\circ}$C, although it can easily adapt to replicate at 30$^{\circ}$C and 43$^{\circ}$C \cite{arribas:2014,somovilla:2019,arribas:2021,somovilla:2022}. For this study, the phage was evolved at three different (30$^{\circ}$C, 37$^{\circ}$C, 43$^{\circ}$C), yet constant each, temperatures starting with a common ancestral population (see Materials and Methods and Fig.~\ref{fig:experiment}B). The evolution temperatures were chosen according to the lab's previous experience \cite{arribas:2014,somovilla:2019,arribas:2021,somovilla:2022} and taking into account that two of them impose strong selective pressures on the virus. At 30$^{\circ}$C, all host metabolic processes are slowed down and, at 43$^{\circ}$C, the heat shock response is fully activated in {\it E. coli}. 

			The fragment from 1060 to 1331 was deep-sequenced in populations at passage 60. Nucleotides from positions 1060 to 1320 code for a fragment of the A2 maturation-lysis protein, 1321 to 1323 correspond to a termination codon and 1324 to 1331 are non-coding positions. The curation of the raw data consisted of removing low-quality sequences not aligned to the reference genome and sequences with insertion and deletion events. This process resulted in high-quality sequences of equal length that were later collapsed into unique sequences while keeping a record of their abundance (see Materials and Methods and Figure~\ref{fig:experiment}C, D). 

			After data curation, we were left with three sets of unique sequences (one for each experimental condition) and their abundances. In the remaining of this paper we only refer to these selected subsets, subsequently used to reconstruct one genotype network for each condition. The three networks for each evolved population are noted $G(T)$, with $T = 30, 37, 43$ labeling the temperature. The aggregated network is noted $G \equiv \cup_T G(T)$. Each node in a genotype network corresponds to a specific sequence in the final sets; two nodes are connected if they differ in a single nucleotide, Figure~\ref{fig:experiment}E. 

			Table~\ref{tab:properties} summarizes the size of each of the final sets, showing the total number $s(T)$ of sequences in the curated sets, the number $N(T)$ of different sequences at each temperature and the total number of links $N^L(T)$. These networks are the largest connected component of the set of sequences in the curated sets. A small fraction of genotypes (below 0,3\% in all cases) not connected through observed genotypes to the connected component are not included in the analysis. 

			Finally, in our representation of genotype networks we depict each node with a size proportional to the logarithm of the abundance $\ln s_i$ of that sequence $i$ (i.e., the number of times that the corresponding sequence was found in our curated datasets, $\sum_i s(i)=s(T)$ for sequences in each of the experiments). As it will be shown, node size directly affects the structure of reconstructed networks. For later convenience, we define the root sequence as the most abundant sequence in each population. 

			\begin{table}[t!]
				\centering
				\begin{tabular}{cccc}
					& $G(30^\circ C)$ & $G(37^\circ C)$ & $G(43^\circ C)$ \\ 
					\hline
					$s(T)$ & 222\,785 & 243\,043 & 315\,912 \\
					$N(T)$ & $9\>188$ & $11\>252$ & $12\>724$ \\ 
					$N^L(T)$ & $19\>531$ & $24\>652$ & $28\>189$ \\ 
					\hline
					$\left<k\right>$ & $4.3 \pm 13.5$ & $4.4 \pm 13.8$ & $4.4 \pm 14.1$ \\ 
					$\alpha$ & $2.63(6)$ & $2.67(7)$ & $2.51(4)$ \\ 
					$\beta$ & $0.73(1)$ & $0.71(2)$ & $0.69(2)$ \\ 
					$r$ & $-0.989$ & $-0.980$ & $-0.970$ \\ 
					 \hline 
					Fixed & G1122A & T1295C & A1088G \\
					mutations & & & G1312A \\
					\hline
				\end{tabular}
    				\caption{{\bf Summary of values for network size and main topological properties.} Total number of sequences $s(T)$ after raw data curation, and number of nodes $N(T)$ and links $N^L(T)$ in the 
				largest connected component of networks reconstructed from our data sets. Lines 4 to 7 show main topological properties: average degree $\langle k \rangle$ and its standard deviation, the exponent of the power-law in the fit to degree abundance ($\alpha$) and to the neighbor connectivity $k_{nn}(k)$ ($\beta$) with the corresponding fitting error between brackets, and Pearson's assortativity coefficient $r$. Last two lines indicate the mutations fixed in the consensus sequence of each population with respect to the wild type; only G1122A is a synonymous mutation.
				\label{tab:properties}}
			\end{table}

  \subsection*{Tetrahedrons, triangles, and squares are the universal building blocks of genotype networks}
    \label{sec:res.1} 

    \begin{figure*}[h]
      \centering
        \includegraphics[width=\textwidth]{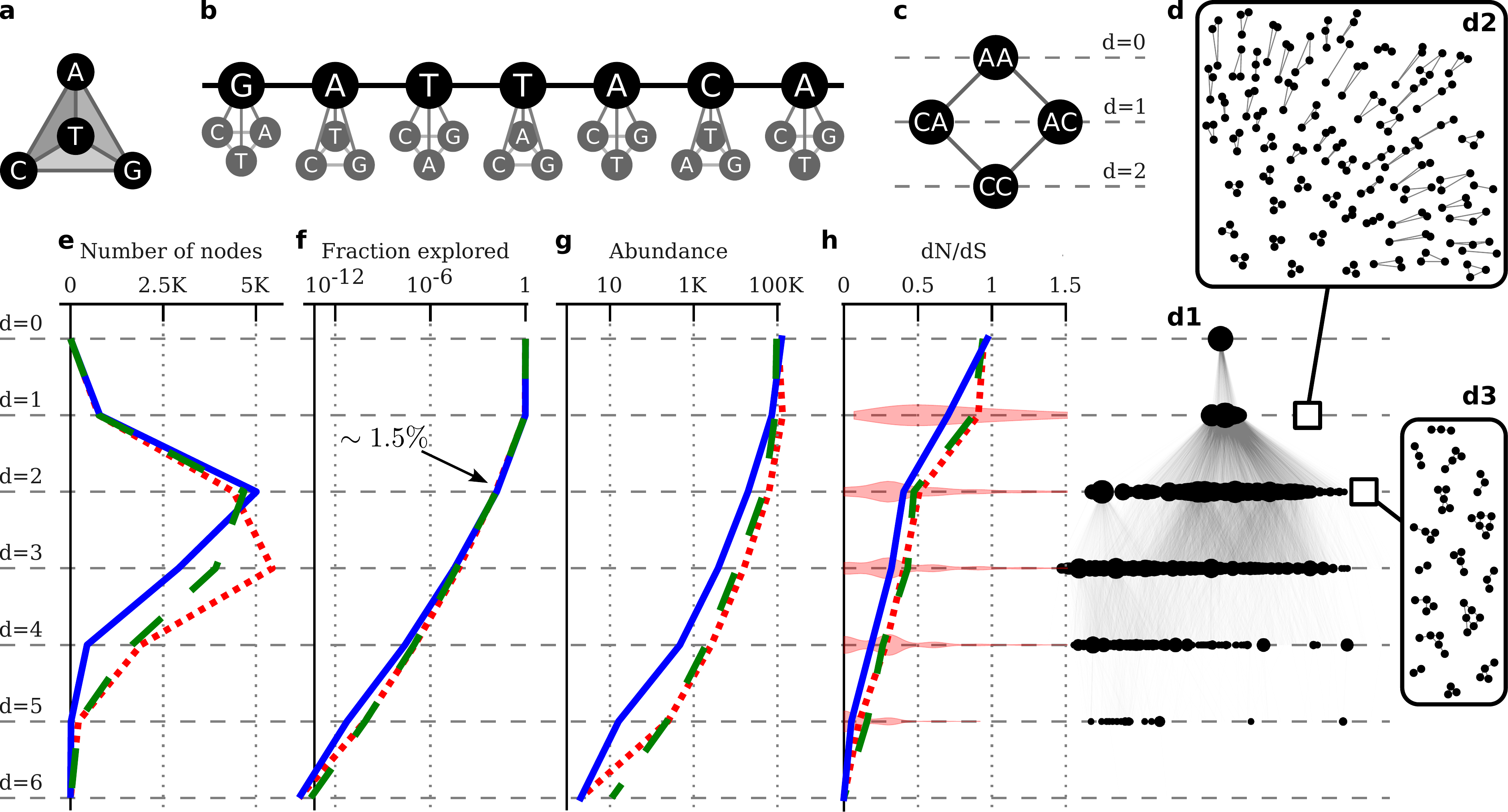}
        \caption{{\bf Microscopic structure of genotype networks and local search.} (a) Basic tetrahedron motif generated by considering all possible values of a specific DNA sequence nucleotide. (b) Necklace structure, with the root sequence as backbone, generated by considering all mutants at distance $d=1$ from the root sequence. (c) Basic square motif generated by considering the two paths leading to a double mutation. (d) Hierarchies emerge in genotype networks of a viral quasispecies as the building motifs are explored and combined. (d1) Hierarchical representation of empirical data from the root ($d=0$) to mutants at each distance found in the deep sequencing of the Q$\beta$-phage populations. Data correspond to network $G(43)$. The ensemble of sequences at distance $d=1$ (d2) and $d=2$ (d3) form simple motifs that, within each level, are disconnected. For genotype networks to be connected, mutants at various distances from the root need to be included. (e-h) Genotype network motifs are exhaustively explored near the root in Q$\beta$ populations; averages at each distance are performed over the ensemble represented in (d1) and over the two remaning populations: blue lines correspond to $T=30^{\circ}$C, green to $T=37^{\circ}$C and red to $T=43^{\circ}$C. (e) Number of unique sequences identified at distance $d$ from the root. (f) Fraction of sequence space explored as a function of distance. (g) Total abundance of sequences sampled at each distance from the root. (h) Ratio of non-silent versus silent mutation fraction at each distance. Violin plots represent the density of $dN/dS$ values at each distance in $G(43)$. Violin plots for $G(30)$ and $G(37)$ are similar, but here omitted for clarity (see Appendix D). 
        }
        \label{fig:hierarchy}
    \end{figure*}

		Basic building blocks (structural motifs) of genotype networks can be derived from the mathematical properties of sequence spaces. Take the root sequence as reference and focus on one of its nucleotides. All possible mutations on that site, while holding the rest of the sequence unchanged, results in $A-1$ mutant sequences at distance $d=1$ from the root, where $A$ is the size of the alphabet ($A=4$ for RNA or DNA sequences). These mutants are connected to the root and are also at distance $d=1$ from each other, thus also mutually connected. This generates a clique or simplex of $A=4$ elements, with the topology of a tetrahedron (Fig.\ \ref{fig:hierarchy}{\bf a}). The same reasoning is valid about any other nucleotide in the root, thus we have a tetrahedron associated to each site along the sequence. All those tetrahedrons share a node, the root, so that the genotype network at distance $1$ looks like a necklace of strung-together $4$-cliques (Fig.\ \ref{fig:hierarchy}{\bf b}): for simplicity, only the mutated nucleotide is shown in the illustration, but each of those mutations is found in a full sequence that shares the composition at all other sites with the root sequence. The tetrahedron simplex is the main motif of genotype graphs. Larger-scale network structures are built by combining tetrahedrons (or parts thereof) around each new sequence. 

		The root acts as a backbone holding together the tetrahedron necklace. If the root is removed, each $4$-element simplex is reduced to a $3$-node clique, forming a second elementary motif: the triangle. These triangles contain all the sequences at distance $d=1$ (of which there are $3L$ different ones). In the absence of the root, they are disconnected from each other, revealing that the set of mutants at distance $d=1$ is massively fragmented. The same is true about the set with all sequences at distance $d=2$. The triangle motif somehow acts as a generator of diversity at fixed distances from the root (see Appendix A). 

		Genotype networks become connected through links between sequences at two different distances from the root. If we focus on mutations in two nucleotide positions, while holding the rest unchanged, we obtain pathways that diverge from the root and meet at the vertex of another set of strung-together tetrahedrons (Fig.\ \ref{fig:hierarchy}{\bf c}). These pathways conform a square, the third relevant motif. Squares involve two changes in distance from the reference sequence, but they result in the integration of the fragmented subsets: While triangle motifs generate all sequence diversity at fixed distances, the resulting subspaces are fragmented; square motifs link them and make them mutually accessible.

	\subsection*{Q$\beta$ quasispecies perform exhaustive local searches in sequence space}

		There are $4^{271} \sim 10^{163}$ different sequences of length $L=271$. The total number of virus particles on Earth is estimated at $10^{31}$ \cite{mushegian:2020}, while natural quasiespecies rarely exceed $10^{12}$ viral genomes within a host. Therefore, natural populations may only sample a tiny fraction of the total sequence space; an even smaller amount (of order $10^6$ sequences) is available to study through deep sequencing methods. Since viral genomes in a natural quasispecies are connected through replication and mutation processes, they typically cluster in sequence space. Our dataset reveals, for each final population, a highly abundant, different sequence (coinciding in the three experiments with the consensus sequence of each population): one or more mutations have been fixed along adaptation to different temperatures. 

		Tetrahedrons, triangles, and squares constitute an abstract scaffolding that underlies every network built from DNA or RNA sequences. This does not imply that all motifs have to appear in empirical networks, as these might be undersampled. As triangles and squares are composed along increasing distances from the root sequence, the connected network of related mutants builds up. Though its basic topology results from the fundamental building blocks described in the previous section, a non-trivial, hierarchical structure emerges as the quasispecies explores nearby mutants (Fig.\ \ref{fig:hierarchy}{\bf d1} and Figure S1). Figure \ref{fig:hierarchy}{\bf d} shows how fragmented groups of genotypes at each distance from the root (Fig.\ \ref{fig:hierarchy}{\bf d2-3}) are held together by paths linking sequences across different levels in the hierarchy. 

		For sequences of length $L$ and an alphabet of $A$ letters the number $S(d)$ of different sequences at a distance $d$ from the root follows 
			\begin{eqnarray}
				S(d) &=& \frac{(A-1)^d L! }{(L-d)! \> d!} \, , 
			\end{eqnarray}
		with an asymptotic behavior $S(d) \sim L^d$ for sufficiently large distances, but far from the length considered, $L \gg d \gg 1$. A comparison with our empirical data allows to estimate the genotype space covering of the populations. Close to the root, coverage is almost complete, and decreases as distance increases: we have found around $95\%$ of all mutants at distance $d=1$ in either population. At $d=2$, $S(2)=329\>265$ for $L=271$. Of these, $5\>013$ (1.5\%), $4\>677$ ($1.4\%$), and $4\>394$ ($1.3\%$) are found in our experiments at $30$, $37$, and $43^\circ C$ respectively. Though empirical sequence diversity peaks at distances $d=2$ or $d=3$ from the root (Fig.\ \ref{fig:hierarchy}{\bf e}), the fraction of sequence space explored decays monotonically, and near to exponentially fast with $d$ (Fig.\ \ref{fig:hierarchy}{\bf f}). 

		The decay in total sequence abundance with distance (Fig.\ \ref{fig:hierarchy}{\bf g}) suggests that mutation plays a role analogous to diffusion in this high-dimensional space. The population is anchored around the root, which could be interpreted as a sequence of high replicative ability (or high fitness, in this context): its copies abundantly cover the neighborhood at distance $d=1$ and, through a cascade-like process, also cover fractions of mutants at higher distances, with a substantial decrease in abundance as the number of mutations increases. Since the number of different sequences grows exponentially fast as $d$ increases, the abundance per node decreases rapidly, and the exhaustive exploration of mutant sequences results severely suppressed a few mutants away from the root: low abundance should correlate with low degree. Figures S4-S6 (Appendix E) represent full networks of $G(T)$ centered at the root and with nodes occupying concentric circles at each distance, thus highlighting their hierarchical structure and qualitatively illustrating the description above. 

		This hierarchical structure is robust and independent of the temperature at which each population has evolved. In particular, there are no obvious signatures of an underlying fitness landscape affecting network structure or preference for certain sequences. In order to quantify the strength and mode of selection in the quasispecies, we have measured the fraction of non-synonymous versus synonymous mutations $q(i|d)=dN(i|d)/dS(i|d)$ in the coding fragment of every sequence, in all three populations. With this notation, we explicitly indicate the distance $d$ at which sequence $i$ is found. First, the set of possible non-synonymous $N(i|d)$ and synonymous $S(i|d)$ mutations for each sequence $i$ has been found. Next, we have counted how many non-synonymous $\tilde{N}(i|d)$ and synonymous $\tilde{S}(i|d)$ mutations are observed in the set of nearest neighbors of sequence $i$ in each sampled population, to obtain 
			\begin{equation}
				q(i|d) = \frac{dN(i|d)}{dS(i|d)} = \frac{\tilde{N}(i|d)/N(i|d)}{\tilde{S}(i|d)/S(i|d)} \, .
			\end{equation}

		Figure \ref{fig:distros}h summarizes the obtained results by showing the average ratio $\bar q_d=\sum_{i|d} q(i|d)$ at each distance $d$ to the root for the three populations, and the density of $q(i|d)$ values in $G(43)$. For the root sequence, with $d=0$, $\bar q_0 \simeq 1$, so its nearest neighbors cover almost all possible mutations, regardless their selective value. No dispersion is shown for $d=0$, since only the root sequence is included. At $d=1$, we observe that an average value $q_1 \lesssim 1$ accompanies the massive coverage of neighboring sequences, as shown by the high dispersion of the distribution of $q(i|1)$ values. In Appendix D, violin plots are shown in full, as well as the the distributions corresponding to populations at $30$ and $37^{\circ}$C, Figure S3. As $d$ increases, the average value $\bar q_d$ decreases towards $0$, with an accompanying diminishing dispersion in $q(i|d)$ values, which progressively cluster around $0$. These results highlight once more the deep correlation between high sequence abundance and exploration of nearby mutants. We interpret that the fitness landscape shows up through purifying selection only at low sequence abundance, limiting in turn the number of deleterious mutants sampled. Altogether, the exploration of variants is restricted to the nearest neighborhood of highly abundant sequences due to the dilution caused by random mutation in the high-dimensional spaces where quasispecies evolve.

	\subsection*{Topological properties are conditioned by the hierarchical structure of Q$\beta$ genotype networks}
    \label{sec:res.2} 

		\begin{figure*}[h]
			\centering
			\includegraphics[width=\textwidth]{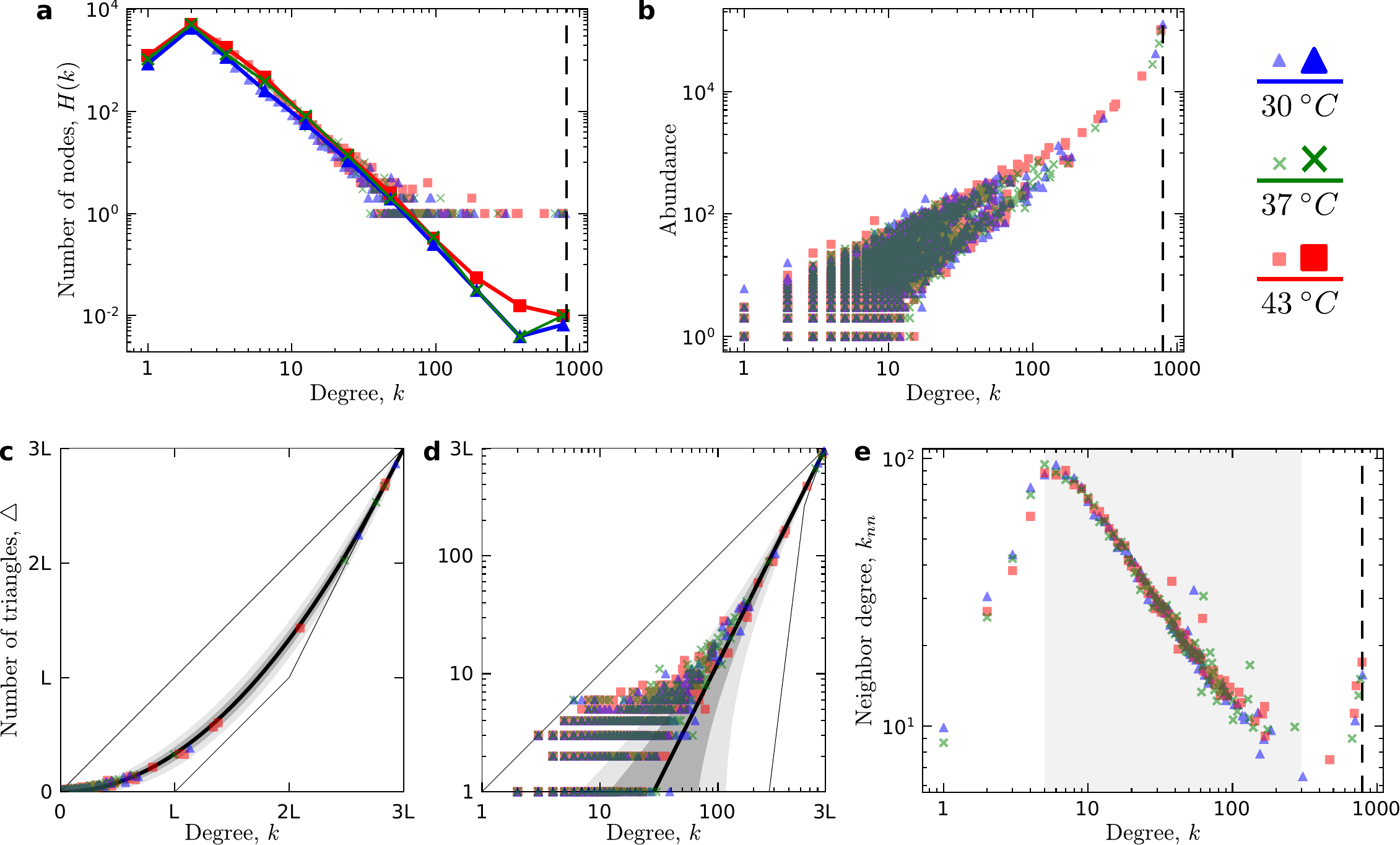}
			\caption{{\bf Macroscopic structure of quasispecies's genotype networks. } {\bf (a)} Degree frequency of genotype networks at $30^{\circ}$C (blue triangles), $37^{\circ}$C (green crosses), and $43^{\circ}$C (red squares). A dashed vertical line marks maximum degree $k_M=3L$ (also in panels {\bf b} and {\bf e}). {\bf (b)} Scatter plot of sequence abundance versus degree. {\bf (c,d)} Number of triangles $\triangle_i$ as a function of the degree $k$ of the sequence in linear {\bf (c)} and logarithmic {\bf (d)} axes. The polygon enclosed by thin continuous lines is defined by maximum $\triangle_M$ and minimum $\triangle_m$ values of $\triangle_i$; the thick continuous line is the expected value $\triangle_{\rm rnd}$ for a random distribution of links, with standard deviation shown as shadings ($\sigma_{\triangle}$, dark grey, and $2 \sigma_{\triangle}$, light grey). {\bf (e)} Network assortativity: average neighbor degree $k_{nn}(i)$ of a sequence's neighbors as a function of sequence degree $k_i$. The left-most part of this plot corresponds to undersampled, peripheral nodes with very low degree, while the right-most part of the plot is dominated by nodes with degree close to saturation, $k_i \simeq 3L$. Only the central part of the plot is statistically representative.}
			\label{fig:distros}
		\end{figure*}

		The three genotype networks reconstructed from each of our experiments display quantitatively identical topological features, despite the different evolutionary history of each population and variations among consensus sequences. Their structural similarity is revealed through an analysis from a complex networks viewpoint \cite{newman:2010}. In this section, we present and discuss three main topological quantities: the degree distribution, the abundance of triangles (related to the clustering coefficient) and the assortativity of reconstructed genotype networks. 

		\paragraph{Degree.} The degree $k_i$ of a node $i$ is defined as the number of links a node has. In our case, it corresponds to the number of sequences in the 1-mutant neighborhood present in the population sample. A histogram of the abundance of nodes with each degree yields a clear signature of a hierarchical organization in the reconstructed networks. Figure~\ref{fig:distros}a shows that all degree-frequency histograms $H(k)$ are well fit by a power-law function of the form $H(k) \propto k^{-\alpha}$ in the interval $2 \le k \le 2L$ (see also Table~\ref{tab:properties}). Nodes with degree $k=1$ are typically undersampled because of their low abundance, while we find an over abundance of nodes with degree close to the maximum possible, $k_{M}=3L$. In these populations, the abundance of a sequence in the sample and the number of neighbors it has (its degree) are positively correlated (Figure~\ref{fig:distros}b): more abundant sequences typically produce more progeny and, under unfaithful replication, populate its immediate neighborhood with higher frequency. 

		The hierarchical structure of these networks also implies that highly abundant sequences (the root and the set of its nearest neighbors, plus some isolated variants at higher distances) have degree higher than average, while the majority of peripheral sequences ($d \ge 4$) are rare and have typically below-average degrees. Actually, the average degree is similar in all three networks, but this value is not representative of a typical node. This is the rule for quantities with power-law-like probability distributions, where the fat tail has a large impact on the average, thus affected by a large standard deviation (see Table \ref{tab:properties}).

		\paragraph{Triangles and clustering.} The local clustering coefficient is a broadly-used measure of the degree of connectivity among neighboring nodes. It is defined as the ratio between the number of triangles formed between a given node and its nearest neighbors and the total number of possible triangles, typically assuming that all nearest neighbors can be mutually connected. But this is not the case in sequence spaces, where a large fraction of triangles cannot be formed, as described above (see Fig.~\ref{fig:hierarchy}b). A more informative quantity of clustering is the absolute number of triangles $\triangle_i$ in the 1-mutant neighborhood of a sequence $i$ (i.e. the numerator in the definition of local clustering coefficient above). 

		Figures~\ref{fig:distros}c and d display $\triangle_i$ as a function of the degree $k_i$ of each sequence. The hierarchical structure of quasispecies networks and the nature of the compounding basic motifs have direct consequences in the number of triangles that can be formed. Consider a specific nucleotide in a given sequence, which can contribute at most a basic tetrahedron motif if the three possible mutations at that site are found in the sample (Fig. \ref{fig:hierarchy}a). If none or only one of the sequences with a mutation at this site are present, this site contributes no triangles to the reference sequence; if two of the mutated sequences are found, then one triangle is formed, and if all three mutations are present, three triangles are formed (the complete tetrahedron). The maximum $\triangle_M$ and minimum $\triangle_m$ number of triangles, defined as the sum of triangles contributed by all sites in a sequence of degree $k$, define a polygon represented in Fig.~\ref{fig:distros}c, d. If mutations are randomly distributed along a sequence, the expected number of triangles is $\triangle_{\rm rnd}=k^2/(3L)$ (thick solid line). The derivation of these functions, as well as the standard deviation $\sigma_{\triangle}$ with respect to the random case is included in Appendix C. Figures \ref{fig:distros}c and 3d show how measured values of $\triangle_i$ compare to $\triangle_M$, $\triangle_{\rm rnd}$ and $\triangle_m$. For degrees $k \gtrsim L$, the empirical value agrees with the random expectation, $\triangle_i (k \ge L) \in \{\triangle_{\rm rnd} - \sigma_{\triangle}, \triangle_{\rm rnd} + \sigma_{\triangle}\}$; one standard deviation corresponds to the dark-grey shadow in Fig. \ref{fig:distros}c, and two standard deviations to the light-grey shadow. However, the number of triangles significantly deviates from random expectations at lower values of the degree, especially for $k \lesssim 60$, with an asymmetric trend to show many more triangles than expected. Note that cases close to $\triangle_M$ or even at this value (there are instances for $k=3$ and $k=6$) entail that mutations preferentially cluster in a few sites of the sequence, while most other sites do not accept any mutation. These deviations from random expectations show a strong preference to mutate in silent positions, avoiding changes in sites that may affect fitness. This is consistent with measures of the $\frac{dN}{dS}$ ratio: purifying selection becomes more prominent in low-abundance, low-degree sequences. 

		\paragraph{Assortativity.} Assortativity quantifies how similar two nodes are with respect to a certain property, in this case node degree. The average neighbor degree, $k_{nn}(i) = {\frac{1}{k_i}} \sum_{j=1}^{k_i} k_j$ measures the average degree of the nearest neighbors of node $i$. Figure~\ref{fig:distros}e represents $k_{nn}(i)$ as a function of $k$ for the three networks studied, showing an inverse dependence between both quantities in the representative region of $k$ values; $k_{nn}(i) \propto k^{-\beta}$, with $\beta > 0$ is a decreasing function of $k$. An additional measure is the assortativity coefficient $-1 \le r \le 1$ (Tab.\ \ref{tab:properties}) (aka the Pearson correlation coefficient of degree between pairs of linked nodes), which measures the correlation between nodes of different degree \cite{newman:2002}. Both quantities show that Q$\beta$ genotype networks are highly disassortative (see Table \ref{tab:properties}), as the neighbors of sequences with high degree preferentially connect to low-degree mutants, while a Pearson coefficient $r \simeq -1$ corresponds to networks that are near complete disassortativity.

	\subsection*{The aggregated genotype network uncovers incipient speciation}
    \label{sec:res.3} 

    \begin{figure*}[h]
      \centering
        \includegraphics[width=0.8\textwidth]{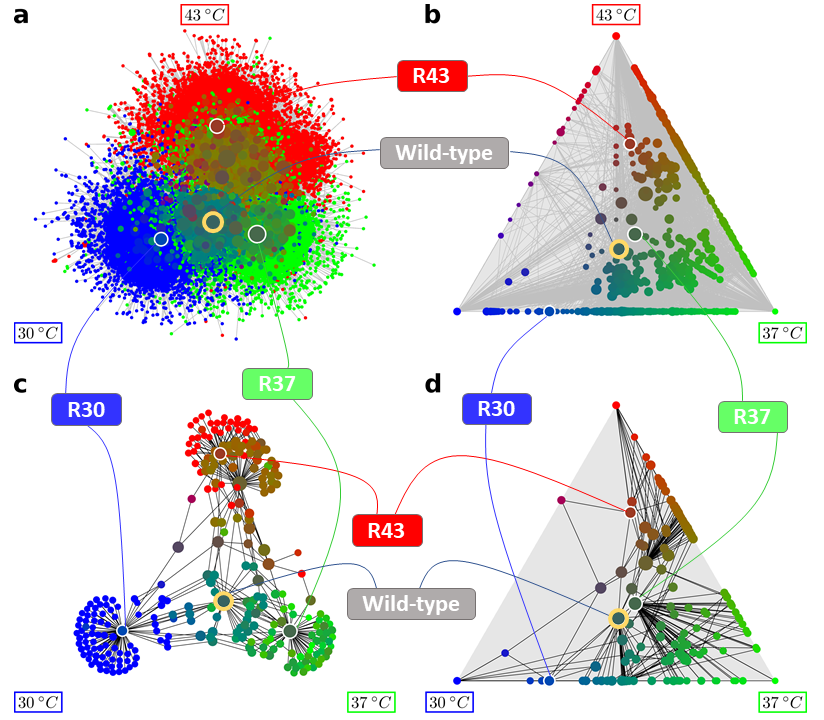}
        \caption{{\bf Visualizing all three genotype networks together.} The color of nodes is weighted proportionally to the sum of the logarithm of the abundances in each network, see Appendix B for details. Pure blue, green or red color correspond to nodes only found in $G(30)$, $G(37)$ or $G(43)$, respectively. Nodes with mixed colors are found in two or more populations. Yellow circles signal the wild-type ancestral sequence, as labeled, and white circles highlight the root sequence of each adapting quasispecies, with labels R30, R37 and R43. {\bf a}. Spring layout representation of all sequences in the aggregated network $G$. {\bf b}. All sequences in a triangle representation. Each experimental condition is mapped in a different vertex. Sequences exclusive of a given environment are collapsed in the corresponding vertex; sequences along edges are found in the two environmental conditions linked by this edge; sequences in the interior of the triangle have been found in all three environments. {\bf c}. Spring layout representation of top $100$ most abundant nodes. {\bf d}. Top $100$ most abundant nodes in triangle representation.}  \label{fig:netTriangle1}
    \end{figure*}

		Each individual genotype network corresponds to a quasispecies adapting to a specific experimental condition. Though the three networks are indistinguishable regarding their topological properties, they are centered at different positions in sequence space, since they have fixed at least one mutation in their root sequence with respect to the wild type (see Table~\ref{tab:properties}), and roots are two or three mutations away from each other. These networks are all part of a larger network, and jointly cover a larger fraction of the genotype space available to Q$\beta$ quasispecies. By examining the aggregated set of sequences in all three experiments, we get a hint of how this larger space is connected and explored by viral populations in response to environmental changes. 

		In Figure~\ref{fig:netTriangle1} two different representations of the complete network $G$, Fig.~\ref{fig:netTriangle1}a, b, and a subset of its most abundant nodes Fig.~\ref{fig:netTriangle1}c, d, illustrate its overall architecture. In all four plots, the wild type sequence and the root sequences of $G(30)$, $G(37)$ and $G(43)$ (labeled R30, R37 and R43, respectively), are highlighted. Fig. \ref{fig:netTriangle1}{\bf a} shows all viral variants within the aggregated network $G$. This representation has been generated in spring layout, a widely used algorithm that treats every link as a spring and determines the position of each node in the image as that which minimizes tension among nodes. In this way, groups of more strongly connected nodes are typically placed closer to each other (and around the center of the illustration), and less connected nodes occupy the periphery. To aid in the visualization of $G(T)$ and to account for specific features of the network (as a sum of sequences in three independent experiments), we have devised an alternative representation in Fig. \ref{fig:netTriangle1}{\bf b}. Here, nodes have been arranged over an equilateral triangle. Each experimental condition ($30^{\circ}$C, $37^{\circ}$C, and $43^{\circ}$C) is associated to a corner and a color (bottom left and blue, bottom right and green, top and red; respectively). Each node's position is a weighted average according to each sequence's abundance under each temperature. Nodes are also colored similarly to their positions, by weighting blue, green, and red (see Appendix B). Only color has been projected into Fig.~\ref{fig:netTriangle1}{\bf a}, while color and position carry the same information in Fig.~\ref{fig:netTriangle1}{\bf b}. In both panels, each node's size is proportional to the logarithm of the corresponding sequence's abundance summed across all three genotype networks. To further clarify the organization of the aggregated network, Figs.~\ref{fig:netTriangle1}{\bf c} and {\bf d} show a subset of $G$ containing only the $100$ most abundant sequences at each experimental condition, all other specifications being the same. 

		Sequences found only at a given temperature appear therefore collapsed at the vertices in Fig. \ref{fig:netTriangle1}b and d, while in Fig. \ref{fig:netTriangle1} c they unfold as a fan of nodes connected to the root at each temperature. Note that if each viral population would contain sequences only adapted to the specific environment where it has evolved along the 60 passages, triangles in Fig. \ref{fig:netTriangle1}b and d would be populated only at the three vertices, and empty elsewhere. Sequences present in samples at two temperatures are found along the edges of the triangle. The least-populated area in the triangle lies near the edge between the $30^{\circ}$C and $43^{\circ}$C corners with a set of sequences populating the edge that apparently work well at both extremes, but not at $37^{\circ}$C. 

		Sequences strictly inside the triangle, which is well populated, show up in all three experimental conditions. Thus, at each temperature, the quasispecies sustains a large variability of genotypes also potentially adapted to other conditions. These versatile sequences appear in notable abundance (large node sizes).  Supporting figure S2 offers a more dynamic view of how the aggregated network is occupied and explored by depicting each network separately. Maintaining node location and color as here described, it informs of each variant across experiments. 

		The joint representation of the three populations adapting to different environments shows how quasispecies start diverging along ecological adaptation. The consensus sequence of the ancestral population $P_{37}^0$ for all three experiments (labeled "wild type") is located near the triangle center. After 60 passages, each quasispecies has incorporated mutations in a root sequence that has dragged the accompanying ensemble towards a different region of sequence space. The wild type is still present in all three populations, and these show a large overlap among variants that are potentially preadapted to different environmental conditions. In physics terms, we witness a process of symmetry breaking, where mutation fixation opens a region of sequence space but limits access to other possibilities, in a drifting process along which the original state becomes progressively forgotten. In the triangle representation, this process can be visualized as a flux from the center to each corner. Still, all possible variants are connected in a single network, suggesting that, at this stage of adaptation, the process is reversible should the environmental conditions vary. 

  \section*{Discussion}
    \label{sec:disc}


		This study provides a detailed reconstruction of the largest genotype network to date for a viral quasispecies, the Q$\beta$ bacteriophage, under different environmental conditions. By analyzing the topological properties of these networks, we observe that the structure of the quasispecies at a fixed time is well described by a highly localized, hierarchical network of interrelated mutants with generic architectural features. This hierarchical organization, emerging from basic structural properties of genotype spaces and (replication-mutation) population dynamics, promotes an exploration of genotype space that is exhaustive but local, clarifying how viral quasispecies maintain a high level of genetic connectivity even as they diverge under different environmental pressures.

		Genotype networks of DNA and RNA sequences, with a four-letter alphabet, are governed at the single- and double-mutant scale by simple motifs: tetrahedrons, triangles, and squares. Some topological properties, as the relationship between the abundance of triangles and the degree of a sequence, are dependent on these architectural properties of genotype networks. Squares, in turn, represent minimal cyclic pathways that make combinatorial novelty accessible from parallel evolutionary routes. The root of the $G(43)$ network is a notable example, since it forms a square with the wild type and the two connecting genotypes. Similar constraints apply to any network built by connecting arbitrary sequences that differ in just one symbol. For example, proteins will have $20$-node simplexes (owing to the $20$-`amino acid' vocabulary) instead of tetrahedrons; but their space at fixed distance will appear similarly fragmented into $19$-node cliques of mutually excluding variants, and these units will become connected by square motifs that allow combinatorial exploration. 

		As the distance to the root increases, the genotype space is progressively less well sampled, loosening the constraints that basic motifs impose on graph structure. Indeed, properties such as the degree distribution and the assortativity of Q$\beta$ genotype networks emerge from population dynamics, which in high-dimensional sequence space is akin to a reaction-diffusion process: more successful, highly abundant sequences, generate as much variation as possible in a process that cascades though hierarchical levels. This results in a correlation between sequence abundance and degree that underlies the hierarchical structure. These processes are fundamental in populations of sequences replicating under a finite mutation rate, such that quasispecies genotype networks are bound to present properties common to generic hierarchical networks \cite{ravasz:2003}, localized in sequence space and with well-defined hierarchical levels \cite{fontoura:2006}. Quasispecies undergo a comprehensive local exploration of genotype space, all measures indicating that the replicative ability (a surrogate of fitness) of nearby mutants is not a major determinant of their presence in the population. In other words, neutral paths do not seem to play a major role in the exploration of genotype spaces near highly abundant sequences. This is in agreement with a previous study that measured the diversity of proteins in the P$_{43}$ population and  concluded that protein sequences have empirical abundances proportional to their phenotypic size, that is, to the theoretical number of sequences that map to each protein, with no signatures of selection found \cite{villanueva:2022}. In the current study, the $dN/dS$ ratio also supports an unselective search of variants, especially near the root. Values $dN/dS>1$ for some mutants are expected within populations, where this ratio cannot be taken as a signature of positive selection \cite{kryazhimskiy:2008}. From a statistical viewpoint, however, the monotonous decrease in the average $dN/dS$ value as hierarchical levels increase suggests that purifying selection acts to prevent the exploration of highly deleterious mutants at low sequence density, preferentially sampling neutral variants \cite{kimura:1977}.


		While analyses of quasispecies structure (or viral intrahost heterogeneity) are limited so far, several studies have addressed the reconstruction and analysis of viral genotype networks from consensus haplotypes. A study of enterovirus diversity from individual infected cells showed qualitative similarities with our study, highlighting the {\it ability of viral populations to remain centered on a slowly moving modal genotype while exploring multiple mutational paths, with branches operating in parallel to reach adaptive genotypes many mutational steps away} \cite{dabilla:2024}. Notably, the authors related the spread of the viral population across mutational landscapes to the use of "tunnels" that permit to traverse fitness valleys. The exhaustive sampling near the root here described, blind to the selective value of mutations, suggests a possible mechanistic origin for stochastic tunneling in evolutionary dynamics \cite{iwasa:2004}.  

		Genotype networks reconstructed from hundreds of sequences of influenza A (H3N2) haemagglutinin abound in cycles (mainly triangles and squares) that reflect convergent evolution and positive selection in the evolutionary dynamics of the pathogen \cite{wagner:2014b}. Remarkably, another study has identified a power-law-like degree distribution in an H3N2 updated genotype network and a positive correlation between sequence degree and abundance in that virus \cite{williams:2022}, suggesting that these features occur across scales. The abundance of cycles and their origin in positive selection has been also described in the genotype networks of many human genes \cite{vahdati:2016}, further supporting that certain topological features of genotype networks transcend the specifics of the model studied. It would be interesting to carry out measures of the number of triangles in networks of consensus haplotypes and see which region of possible triangle values they occupy. We believe that the total number of triangles formed among the neighbors of a sequence is a more informative measure than the clustering coefficient, for various reasons. First, it captures the microscopic structure of genotype networks, based on triangles and their simplexes. Secondly, bounds to the number of triangles and the expectation under no selection can be analytically calculated, providing a straight interpretation of the results. Third, it allows a comparison across networks, and may highlight different types of selection acting on dissimilar model systems, at different levels of resolution or at various temporal scales.  

		In contrast with the topological measures above, the disassortativity measured in Q$\beta$ is not generic. Measures in other empirical and synthetic genotype networks cover a broad range of assortativity values, from  positive to highly negative. Genotype networks obtained for the sequence-to-secondary-structure RNA map \cite{aguirre:2011}, yield assortative networks with $\beta \simeq 0.75$ and a Pearson's correlation coefficient $r > 0$, while exhaustively sampled empirical networks of transcription factor binding sites of length 8 have positive assortativity, with $r = 0.25$ \cite{aguilar:2018}. In these two cases, the value of $r$ positively correlates with the size of the network. In one H3N2 network, assortativity was negative, with $r \simeq -0.15$ \cite{williams:2022}. For these apparently dissimilar results to be reconciled, we need to recall that empirical networks are limited by searches performed by natural populations highly localized on vast, hyperastronomically large genotype networks \cite{louis:2016}, of which they only explore a tiny fraction. Networks of transcription factor binding sites are an exception, since the short length of the sequences involved permits an exhaustive exploration of genotypes. In fact, genotype-to-phenotype models support that high-degree sequences tend to cluster together, so full genotype networks are expected to be assortative \cite{bornberg-bauer:1999,wuchty:1999}. In contrast, when the genotype network is minimally sampled, as it is our case, a local hierarchical structure with negative assortativity emerges. As the fraction of network genotypes sampled increases, we hypothesize that assortativity should take larger values, eventually converging to the (positive) value characteristic of the corresponding full network. An improved sampling can be attained through the aggregation of partly overlapping, local hierarchical genotype networks: as a proof of concept, we have measured the assortativity of the aggregated network $G$ and obtained a value $\beta=-0.57(1)$, in support of our hypothesis. 

		A representation of large molecular populations in the form of genotype networks proves to be a valuable method to visualize and quantify features that significantly modify how the overall evolutionary process is conceptualized. While phylogenetic visualizations emphasize irreversible branching, a genotype network approach portrays a more dynamic and fluid picture of evolution, where mutational pathways remain connected, and populations test variants that can be reverted, especially in the early stages of divergence. Genotype networks permit a clear visualization of incipient ecological speciation in Q$\beta$ phage populations. Although these populations are beginning to diverge, they remain connected through mutational pathways and multiple shared genotypes. A large fraction of sequences identified in the evolved populations were already present in the ancestral population, supporting that Q$\beta$ populations sustain a high degree of pre-adaptation \cite{somovilla:2022}. This feature might be also generic in heterogeneous populations, as suggested by a study where genotype networks for two RNA molecules catalyzing two different chemical reactions showed extensive functional overlap promoting evolutionary innovation \cite{bendixsen:2019}. 

		Longer experiments, or experiments that impose more unrelated conditions, should deepen the separation between viral populations, forcing the fixation of additional mutations and turning a reversible process into an irreversible one at larger temporal scales. Divergence causes the progressive loss of memory of the initial condition, resulting in less shared variants and, in our picture, depopulating the inside of the triangle representation. However, the fact that cycles in genotype networks are pervasive also at larger spatio-temporal scales \cite{wagner:2014b,vahdati:2016,williams:2022,dabilla:2024} suggests that representations in terms of genotype networks are needed to capture features of the viral evolutionary process instrumental in its interpretation. Network-based views of viral evolution provide a powerful framework for understanding the fine-scale processes that drive adaptation and divergence in mutant swarms, highlighting at once the existence of pre-adaptation to various environments, the non-directional exploration of sequence spaces along adaptation and the reversibility of the search process.

	\section*{Matherials and methods}

    \subsection*{Experimental adaptation of the Q$\beta$ phage to different temperatures}
			\label{sec:meth.1}

			Plasmid pBRT7Q$\beta$ \cite{taniguchi:1978}, which contained the bacteriophage Q$\beta$ cDNA cloned in the plasmid pBR322, was used to transform {\it Escherichia coli} DH5-$\alpha$. An overnight culture supernatant from a transformed colony was used to infect {\it E. coli} Hfr (Hayes) on semi-solid agar at a multiplicity of infection (moi) that allowed the generation of well-separated lysis plaques (biological clones). The viral progeny contained in a randomly chosen lysis plate was extracted as described \cite{arribas:2021}, subjected to Sanger sequencing \cite{laguna-castro:2022}, and used to infect an exponential-phase {\it E. coli} Hfr (Hayes) culture using an moi of 0.1 in a final volume of 1 mL of NB medium (8 g/L Nutrient Broth from Merck and 5 g/L NaCl). After 2 h of incubation at 37$^{\circ}$C, the culture was treated with 1/20 volumes of chloroform, incubated in a thermoblock for 15 min at 37$^{\circ}$C and 600 rpm, and centrifuged for 10 min at 12 000 rpm to collect the supernatant with the virus particles corresponding to the first passage. A fraction of this supernatant was titrated and used to infect a new {\it E. coli} Hfr culture at 37$^{\circ}$C with an moi around 0.1. Virus population obtained at passage number 2 has been extensively described \cite{somovilla:2022} and was the ancestor $P_{37}^0$ of all the evolutionary lines analyzed in this work (as depicted in Figure \ref{fig:experiment}). The consensus sequence of $P_{37}^0$ has no mutations with respect to the cDNA cloned in pBRT7Q$\beta$. Evolution of the ancestral population $P_{37}^0$ at 30$^{\circ}$C, 37$^{\circ}$C and 43$^{\circ}$C took place for 60 serial passages, carried out under the same conditions above described with the only difference of the incubation temperature \cite{somovilla:2019}. Three replicate lines were established for each temperature although only the passage number 60 of one of them was subjected to deep sequencing. 

    \subsection*{Deep sequencing analysis}
      \label{sec:meth.15}

			Phage RNA was extracted and purified using the Mini Kit Viral RNA QIAamp (Qiagen). It was amplified by RT-PCR using SuperScript II Reverse Transcriptase (Invitrogen) and Q5 Hot Star High fidelity (New England Biolabs), using primers 5'GAATGTTGGTGACATACTTGCT3' (forward) and 5'TTGCCATGATCAAATTGACC3' (reverse). PCR products were quantified and tested for quality (TapeStation 4200, Agilent Technologies), prior to Illumina Ultra Deep Sequencing analysis (MiSeq platform with $2 \times 250$ bp mode and v2 chemistry). After removing the sequence of the primers, the amplicon obtained covered from nucleotide position 1060 to 1331.

    \subsection*{Data preprocessing and curation}
      \label{sec:meth.2}

			The following pipeline was applied to the three datasets P$_{30}$, P$_{37}$, and P$_{43}$. The Illumina MiSeq Next Generation Sequencer produced paired-end reads first subjected to quality control with FastQC v0.11.9 \cite{andrews:2012} and MultiQC v1.9 \cite{ewels:2016}. Because only base pairs near the 3'-end of sequences had a low Phred Quality Score, we used Cutadapt v3.0 \cite{martin:2011} to remove them using a minimum threshold of 25. We also removed primer and barcode sequences with Cutadapt. Reads that could not be adequately trimmed or whose length was smaller than 50 bp were discarded. Since forward and reverse reads overlapped over multiple base pairs, we used Flash2 v2.2.0 \cite{magoc:2011} to merge paired-end reads and obtain a set of high-quality single-end reads. In the next step, we aligned the reads to the Q$\beta$ reference genome with BWA-MEM aligner v0.7.17 \cite{li:2013} and discarded reads that could not be aligned using SAMtools v1.11 \cite{li:2009}. From the resulting BAM files, we further discarded reads containing insertions or deletions or whose length differed from the sequenced amplicon. SAMtools was then used to convert BAM files into FASTA, which contained multiple repeated reads. Therefore, we used Collapser from the \href{http://hannonlab.cshl.edu/fastx_toolkit/}{FASTX-Toolkit} v0.0.14 to collapse repeated reads while keeping a record of their abundance. The resulting FASTA files containing unique reads were used for all subsequent analyses. The pipeline was implemented with a custom Bash script that can be found in our GitHub repository, \url{https://github.com/HSecaira/Quasispecies\_TFM/tree/main/Preprocessing}.










\appendix 
\clearpage

  \setcounter{figure}{0}
  \setcounter{table}{0}
  \renewcommand{\figurename}{SUP. FIG.}
  \renewcommand{\tablename}{SUP. TAB.}

	\section{Fragmented subnetworks at fixed distance $2$ from an arbitrary reference}  
    \label{app:1} 

    In the main text we saw how subnetworks at one point-mutation distance of an arbitrary reference appear as fragmented sets of triangles. Subnetworks at any moderate fixed distance from an arbitrary reference appear heavily fragmented as well, and triangles are again a generating motif of such subnetworks. The role of triangles comes from the alphabet size, which is $A=4$ for RNA and DNA genotype networks. Networks generated from sequences with an alphabet of size $A$ would see $A-1$ simplexes instead of triangles. 
    
\begin{figure}[h]
      \begin{center} 
        \includegraphics[width=\columnwidth]{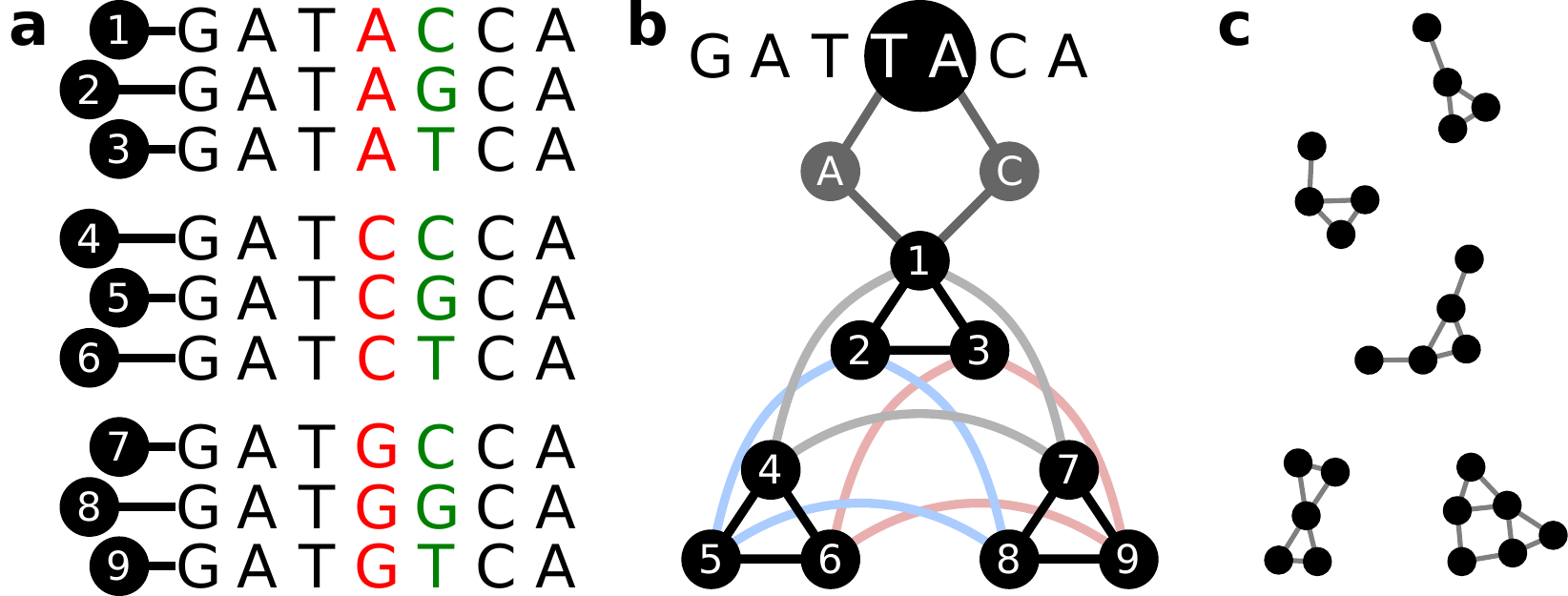}    
        \caption{{\bf Fragmented networks at distance $2$.} }
        \label{fig:distance2}
      \end{center}
    \end{figure}

    Sup.\ Fig.\ \ref{fig:distance2} illustrates network fragments at fixed distance $2$ from a reference. Taking GATTACA as our string reference, Sup.\ Fig.\ \ref{fig:distance2}{\bf a} lists all sequences that are at distance $2$ with mutations in positions $4$ and $5$. These sequences come in triplets which are at distance $1$ from each other, thus all of them appear connected. This is visualized as triangles with solid black edges in Sup.\ Fig.\ \ref{fig:distance2}{\bf b}. Additionally, the $n$-th member of each triplet is at distance $1$ from the $n$-th members of the other two triplets, thus they are also connected in the genotype network. This is illustrated in Sup.\ Fig.\ \ref{fig:distance2}{\bf b} by the arching edges of lighter gray, blue, and red. Links in a darker shade of gray show how each of the nodes in this graph is connected to the reference string by a square motif. 

    The resulting graph reminds us of an external product of a triangle with itself. A similar subnetwork is rendered if we explore all mutations in any other two positions of the reference string. Critically, two such subnetworks are not connected to each other. Take all mutations at distance $2$ from the reference in positions $4$ and $5$ (as displayed in the figure) and all mutations at distance $2$ from the reference in positions $4$ and $6$. While such two subsets of strings could easily have the same nucleotide in position $4$, they will always be different in positions $5$ and $6$, thus they will be at distance $2$ from each other. Hence, given a reference string and two positions, we obtain a subnetwork as that in Sup.\ Fig.\ \ref{fig:distance2}{\bf b}. For any other two positions we obtain a similar graph, but such graphs are disconnected from each other. 

    As it happened with the simpler networks at distance $1$ from a reference, mutations within the current graph stand for substitutions of a given nucleotide. A viral quasispecies navigating subnetworks like this is changing pieces at fixed positions, but not integrating novelty in a combinatorial fashion. That was rather done by the square motif, which combined mutations at two different positions. And that would be made by additional square motifs that move us to subnetworks at distance $3$ and beyond. 

    In our deep sequencing samples we found about $~1,5\%$ of the subspace at distance $2$ from each reference. Hence finding the complete graph from Sup.\ Fig.\ \ref{fig:distance2}{\bf b} is unlikely. Sup.\ Fig.\ \ref{fig:distance2}{\bf c} shows examples of actual graphs found by the deep sequencing sampling---also shown in Fig.\ 2{\bf d3}. All of them must be subgraphs of the network in Sup.\ Fig.\ \ref{fig:distance2}{\bf b}. Note that all nodes within each independent connected component shown in Sup.\ Fig.\ \ref{fig:distance2}{\bf c} must correspond to two different mutations in the same fixed two positions of the reference sequence. While the real quasispecies does not sample the comple subnetwork at fixed distance two, the explored space inevitably inherits the underlying fragmented nature at each fixed distance.

  \section{Mathematical details of triangle plots}  
    \label{app:2} 

    Take an equilateral triangle of unit side with its center in the origin of coordinates $\overrightarrow{0} = (0,0)$. From bottom left, and in anti-clockwise order, the corners of this triangle are located at $\overrightarrow{v}_{30} = (-1/2, -\sqrt{3}/2-1/4)$, $\overrightarrow{v}_{37} = (-1/2, \sqrt{3}/2+1/4)$, and $\overrightarrow{v}_{43} = (0, 1/2)$. These vectors have been labeled according to our three experimental conditions. 

    \begin{figure*} 
      \begin{center} 
        \includegraphics[width=0.9\textwidth]{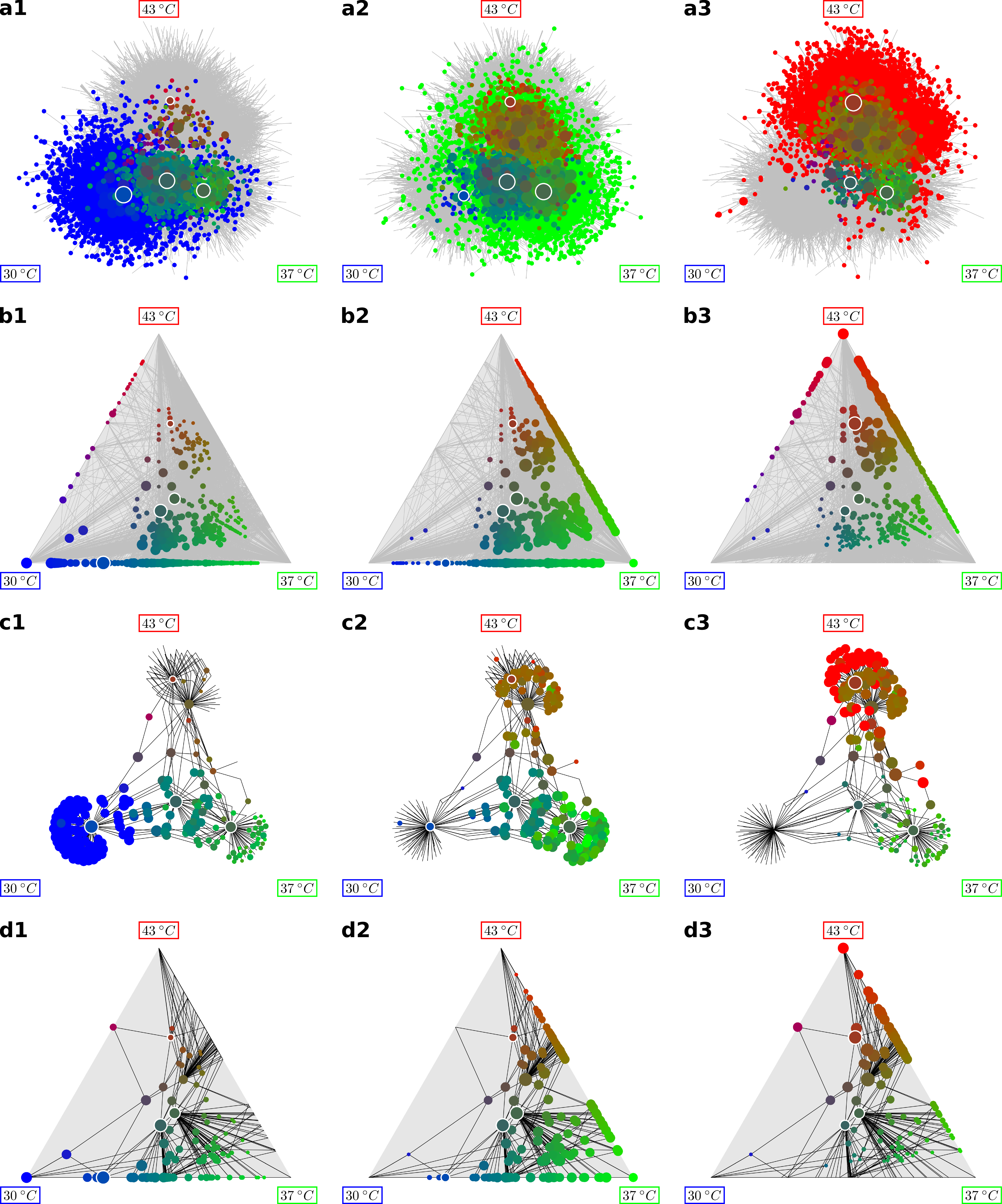}
        \caption{{\bf Visualizing all three networks together.} {\bf a} All nodes in triangle. {\bf b} All nodes in spring layout. {\bf c} Top $100$ nodes in triangle. {\bf d} Top $100$ nodes in spring layout. {\bf a1}, {\bf b1}, {\bf c1}, {\bf d1} Node size is proportional to the logarithm of the abundance of each sequence in the network at $30^\circ C$. {\bf a2}, {\bf b2}, {\bf c2} Node size is proportional to the logarithm of the abundance of each sequence in the network at $37^\circ C$. {\bf a3}, {\bf b3}, {\bf c3} Node size is proportional to the logarithm of the abundance of each sequence in the network at $43^\circ C$.}
        \label{fig:netTriangle2}
      \end{center}
    \end{figure*}

    Take the set of all unique sequences in the grand network: $\Sigma \equiv \{\Sigma_i,\> i=1, \dots, N\}$. Each of these sequences has an abundance array: 
      \begin{eqnarray}
        \bar{a}_i \equiv \left(a_i(30), a_i(37), a_i(43) \right), 
        \label{eq:app.1}
      \end{eqnarray}
    which consists of the number of counts of this sequence in each of the experiments. We want to use these abundances to locate each sequence's node within the equilateral triangle, biasing their positions towards each of the corners depending on each sequence's abundance in each of the experiments. We want to use that information also to determine the size and color of each node. Differences in abundances between different sequences are very large (spaning orders of magnitude), which would lead to very bad representations, as nodes would cram into small regions and huge hubs would dominate the representation. Better visualizations are obtained by basing our representations on logarithms of the abundances. Let us define a log-abundance array for each sequence as: 
      \begin{eqnarray}
        \bar{\alpha}_i &=& \left( \log\left(a_i(30)+1\right), \log\left(a_i(37)+1\right), \log\left(a_i(43)+1\right) \right) \nonumber \\ 
        &\equiv& \left(\alpha_i(30), \alpha_i(37), \alpha_i(43) \right). 
      \end{eqnarray}
    Note that we have added $1$ to each abundance to avoid issues with sequences that are absent in some of the experimental conditions. 

    From here we derive a total log-abundance and a weight array: 
      \begin{eqnarray}
        \alpha_i &=& \alpha_i(30) + \alpha_i(37) + \alpha_i(43), \nonumber \\ 
        \bar{\omega}_i &=& \left(\alpha_i(30)/\alpha_i, \alpha_i(37)/\alpha_i, \alpha_i(43)/\alpha_i \right) \nonumber \\ 
        &\equiv& \left(\omega_i(30), \omega_i(37), \omega_i(43) \right). 
        \label{eq:app.2}
      \end{eqnarray}

    In Figs.\ \ref{fig:netTriangle1}{\bf a}, \ref{fig:netTriangle1}{\bf c}, \ref{fig:netTriangle2}{\bf a}, and \ref{fig:netTriangle2}{\bf c}, the position of each node with the triangle is given by the vector: 
      \begin{eqnarray}
        \overrightarrow{x}_i &=& \omega_i(30)\overrightarrow{v}_{30} + \omega_i(37)\overrightarrow{v}_{37} + \omega_i(43)\overrightarrow{v}_{43}. 
        \label{eq:app.3}
      \end{eqnarray}
    The positions of each node in Figs.\ 4{\bf b}, 4{\bf d}, \ref{fig:netTriangle2}{\bf b}, and \ref{fig:netTriangle2}{\bf d} are determined generating a spring layout with Python's NetworkX library \cite{springLayout}. 
    The layout is the same in Figs.\ 4{\bf b} and \ref{fig:netTriangle2}{\bf b}, and in Figs.\ 4{\bf d} and \ref{fig:netTriangle2}{\bf d}. 

    The colors in all panels of Figs.\ 4 and \ref{fig:netTriangle2} are similarly taken as a weighted sum of red, green, and blue. Taking the color vectors $\overrightarrow{r} \equiv (1,0,0)$, $\overrightarrow{g} \equiv (0,1,0)$, and $\overrightarrow{r} \equiv (0,0,1)$ for each color respectively, a node's color is given by: 
      \begin{eqnarray}
        \overrightarrow{c}_i &=& \omega_i(30)\overrightarrow{r} + \omega_i(37)\overrightarrow{g} + \omega_i(43)\overrightarrow{b}. 
        \label{eq:app.4}
      \end{eqnarray}

    The size of a node in Fig.\ 4 is proportional to its total log-abundance, $s_i \sim \alpha_i$; while the size of a node in each column of Fig.\ \ref{fig:netTriangle2} is respectively proportional to the log-abundances in each of the experimental conditions: $s_i(30) \sim \alpha_i(30)$, $s_i(37) \sim \alpha_i(37)$, and $s_i(43) \sim \alpha_i(43)$.

	\section{Bounds to number of triangles per sequence}
		Take a node of degree $k$ in the genotype network (a sequence with $k$ mutants at distance $d=1$ present in the sample). Let us first estimate the minimum number of triangles $\triangle_m (k)$ that can be formed with $k$ neighbours. If $k \le L$, mutants could correspond all to different nucleotides in the sequence, such that no triangles are formed. For $L < k \le 2L$, the number of triangles is minimized if, once every nucleotide position contributes a link (a mutant), a second link is added to each position; the number of triangles increases at a rate of one triangle per new link. Finally, when all positions have two links, the addition of new links for a total degree $2L < k \le 3L$ contributes two new triangles per link added. Therefore, the minimum number of triangles that a node of degree $k$ can have is a pair-wise continuous function $\triangle_m (k)$ of the form
			\begin{equation}
				\triangle_m(k) = \left \{
				\begin{array}{lc}
				0 \, , & 0 \le k \le L \\
				k-L \, , & L \le k \le 2L \\
				2k - 3L \, , & 2L \le k \le 3L
				\end{array}
				\right.
			\end{equation}
		The maximum possible number of triangles $\triangle_M (k)$ that a node of degree $k$ can have is achieved by filling sequentially each position with the three possible mutants before adding links (neighboring mutants) to other positions. At each position, no triangles are formed with 0 or 1 link, one triangle appears for $k=2$ and three triangles for $k=3$. This is also a pairwise continuous function as $k$ increases that can be written as 
			\begin{equation}
				\triangle_M(k) = \left \{
				\begin{array}{lc}
				3n \, , & k = 3n+1 \\
				3n+1 \, , & k = 3n+2 \\
				3n+3 \, , & k = 3n+3
				\end{array}
				\right.
			\end{equation}
		for $n$ integer, $n \in \{0, 1, 2, \dots (L-1) \}$. Note that the line $\triangle_M(k) = k$, as drawn in Figure 3c and 3d is a global upper bound to the stair function above.

		The number $\triangle_{\rm rnd}(k)$ of triangles expected in a sequence with $k$ neighbors at distance 1, if these links would correspond to random mutations along the sequence, can be approximated as follows. First, the probability that a specific position has a link is $p=k/3L$. The probability that it has a total of $0 \le k \le 3$ links is a binomial distribution $P(n,3,k/3L)$, with $n=0, 1, 2, 3$,
    	\begin{equation}
        p(n,3,k/3L) = {3 \choose n} p^n (1-p)^{3-n} \, .
        \label{eq:pLinks}
    	\end{equation}
		The total number of triangles in a sequence of length $L$ is obtained by adding the contributions of each position: there are $L p(n,3,k/3L)$ positions with $n=0, 1, 2$ or $3$ links, and they contribute $0, 0, 1$ and $3$ triangles, respectively. Therefore,
    	\begin{equation}
        \triangle_{\rm rnd}(k) = L p(2,3,k/3L) + 3L p(3,3,k/3L) \, ,
    	\end{equation}
		yielding
    	\begin{equation}
        \triangle_{\rm rnd}(k) = \frac{k^2}{3L} \, .
    	\end{equation}

		The standard deviation of this value can be also calculated from
    	\begin{equation}
        Var [\triangle_{\rm rnd}] = \sum_{j=0}^3 p(j,3,k/3L) \left( x_j - \triangle_{\rm rnd}(k) \right)^2 \, 
    	\end{equation}
		where $x_j$ is the number of triangles observed for each value of $j$, $x_0=x_1=0$, $x_2=1$ and $x_3=3$. After some algebra, we get 
    	\begin{eqnarray}
        \sigma_{\triangle_{\rm rnd}}&=& \sqrt{Var [\triangle_{\rm rnd}]} \nonumber \\
        &&2\sqrt{L} \times \nonumber \\
        &&\sqrt{3y^2(-3y^5 + 3y^4 + 6y^3 -9y^2 - y + 4)} , \nonumber \\ 
        &&
    	\end{eqnarray}
		with $y\equiv k/3L$. Note that these calculations assume that the probability of having mutations at a specific site are independent from each other. This is true in general, but only approximated if the number of mutations (links) is fixed. For an exact calculation, the $P(n,3,k/3L)$ should be coupled across sites, but this makes calculations much more cumbersome, while corrections are minimal. 

	\section{Statistics of $dN/dS$ values}

		The ratio of non-synonymous versus synonymous mutations has been calculated for each sequence considering as observed mutations those present in the ensemble of their nearest neighbors (one mutation away). Figure S3 shows averages and violin plots for the density of $dN/dS$ values for each experiment and at each distance from the root sequence. As described in the main text, average values monotonically decrease with $d$, from $\bar q_0 \simeq 1$ to $\bar q_6 \simeq 0$. At any distance $d \ge 1$ the dispersion is very large, due to the massive exploration of mutant sequences near highly abundant nodes. Consistently, the exploration of mutants at distances $d > 6$ is severely suppressed in general due to low sequence abundance. 

    \begin{figure*}[h]
      \begin{center} 
        \includegraphics[width=\textwidth]{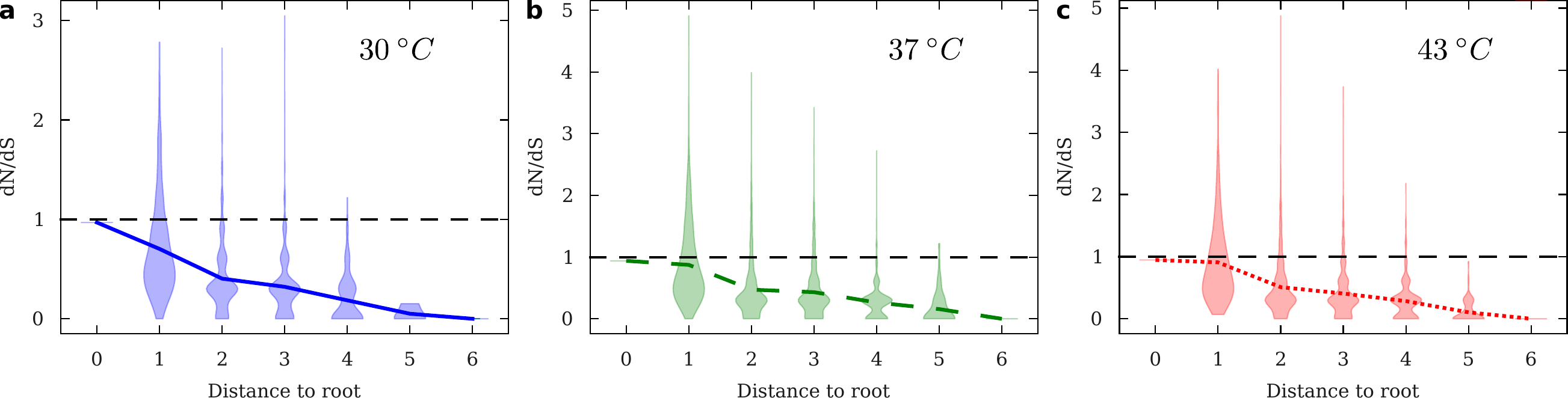}    
        \caption{{\bf $dN/dS$ ratio for networks at each temperature.} Note the difference in the vertical scale. The dashed line at $dN/dS=1$ signals no selection. The three populations show comparable patterns of variation in average and dispersion values, despite their different evolutionary histories.}
        \label{fig:dNdS}
      \end{center}
    \end{figure*}

	\section{Visualization of quasispecies as a hierarchy}

		Figures S2-S4 depict the three genotype networks reconstructed as a hierarchy. The most abundant haplotype is at the center of the network, and concentric circles around it contain the sets of sequences at distances $d=1, 2, 3 \dots$. The average decrease in abundance (in node size) is visible in this representation, qualitatively illustrating the analogy of replication and mutation with a reaction-diffusion process. Most abundant nodes have well-sampled neighborhoods, while rare sequences occupy areas of low diversity (and have low degree). The appearance of possible advantageous mutants at distances $d=1$ and $d=2$ is indicated by nodes with size comparable to the root. The case of $G(43)$ is particularly interesting, with peripheral nodes of high abundance that, for this reason, explore variants nearby in a repetition of the hierarchical structure. Some of these highly abundant nodes might become the new root along evolution and drag the mutant swarm towards its neighborhood.

		This hierarchical structure was depicted using the radial layout from yfiles \cite{wiese:2004} implemented in Cytoscape v3.10.2 \cite{shannon:2003}. For this, we first calculated the distances of all sequences to the root and then applied the radial layout algorithm, which used such distances to organize circles around the root. We also adjusted the size and color of nodes to be proportional to the natural logarithm of their abundance.    

    \begin{figure*}[h]
      \begin{center} 
        \includegraphics[width=0.9\textwidth]{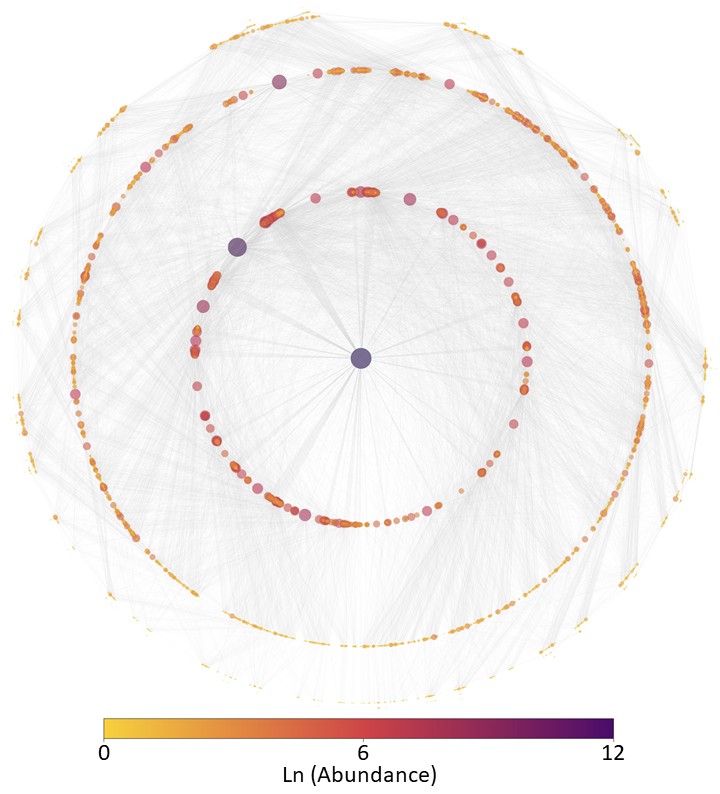}    
       \caption{{\bf Complete genotype network $G(30)$.} Color scale indicates the natural logarithm of the abundance of each sequence (node), which varies from a single sequence to about 160\,000 sequences in the root.} 
        \label{fig:GN30}
      \end{center}
    \end{figure*}

    \begin{figure*}[h]
      \begin{center} 
        \includegraphics[width=0.9\textwidth]{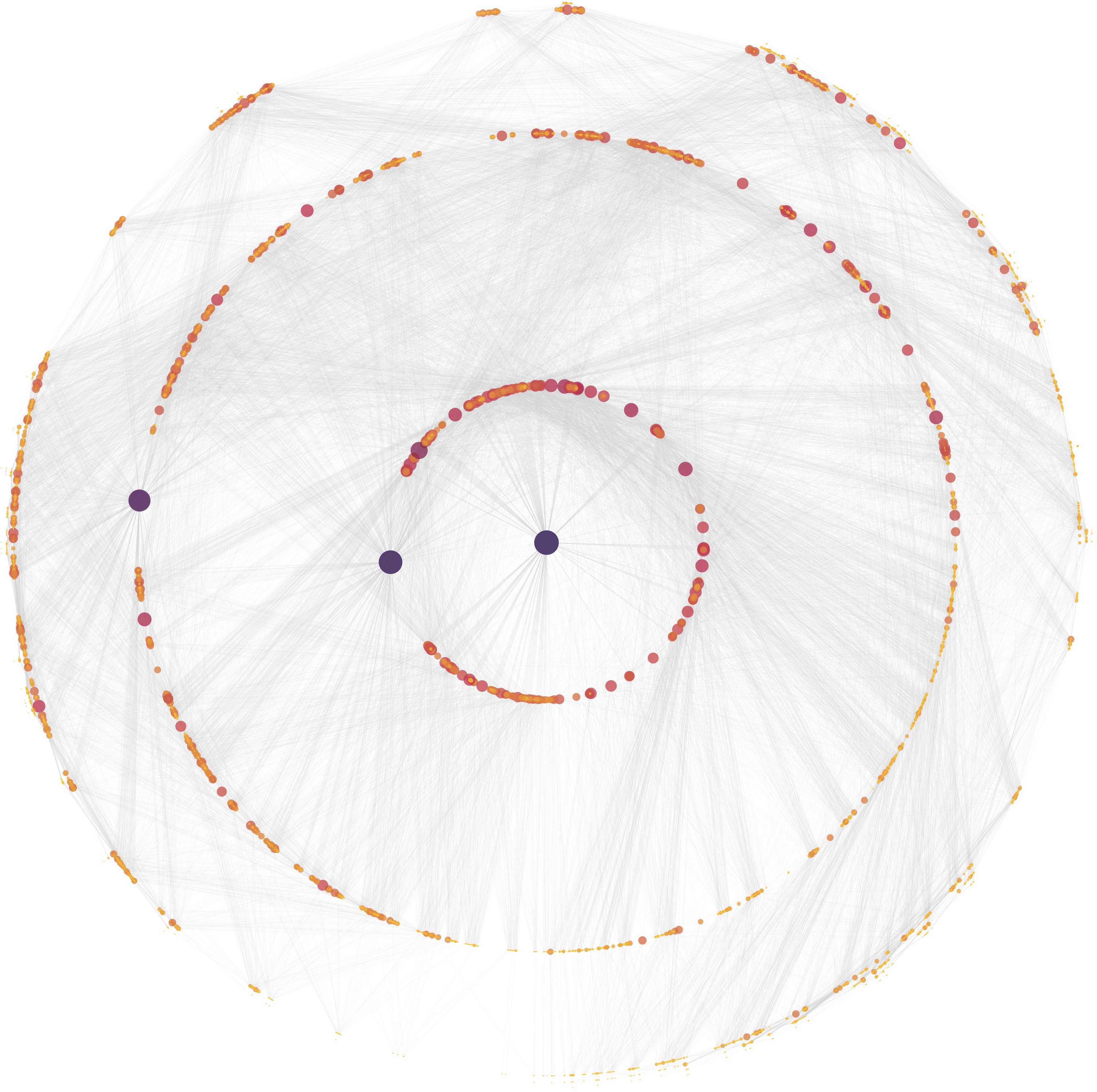}    
       \caption{{\bf Complete genotype network $G(37)$.} Color scale as in $G(30)$.} 
        \label{fig:GN37}
      \end{center}
    \end{figure*}

    \begin{figure*}[h]
      \begin{center} 
        \includegraphics[width=0.9\textwidth]{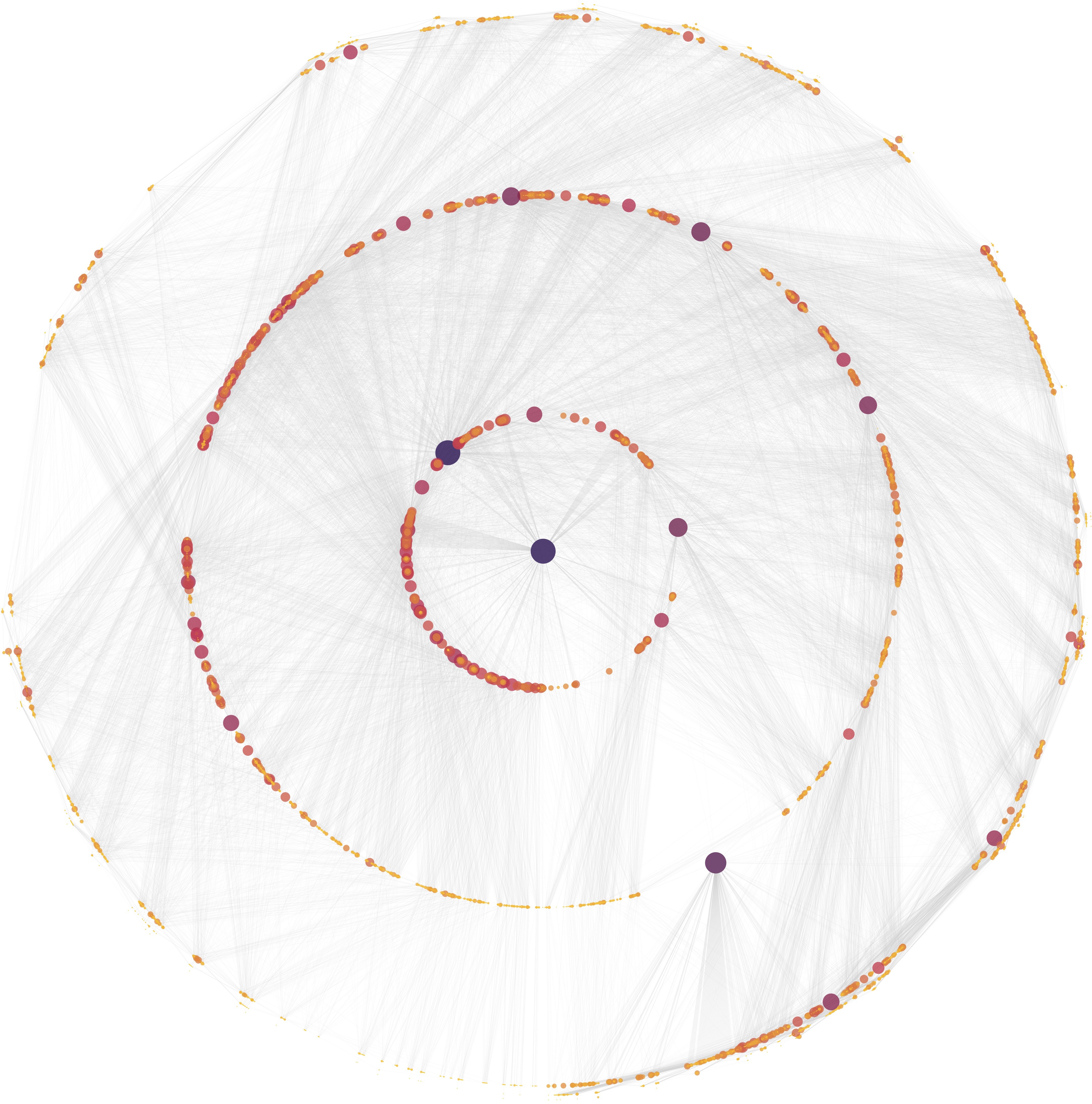}    
       \caption{{\bf Complete genotype network $G(43)$.} Color scale as in $G(30)$.} 
        \label{fig:GN43}
      \end{center}
    \end{figure*}

\end{document}